\documentclass[aps,twocolumn,showpacs]{revtex4}
\usepackage{graphicx}
\usepackage{bm}
\usepackage{graphicx}
\usepackage{amsmath}
\usepackage{mathrsfs}
\usepackage{ulem}
\usepackage{color}
\usepackage[colorlinks, linkcolor=red,anchorcolor=green,citecolor=blue]{hyperref}
\graphicspath{{./figs/}}

\begin{document}
\title{Charmonium transition in electromagnetic and rotational fields}
\author{Shile Chen}
\author{Jiaxing Zhao}
\email{zhao-jx15@tsinghua.org.cn}
\author{Pengfei Zhuang}
\email{zhuangpf@mail.tsinghua.edu.cn}
\affiliation{Physics Department, Tsinghua University, Beijing 100084, China}
\date{\today}

\begin{abstract}
We study charmonia in electromagnetic and rotational fields in the frame of a potential model. Different from the temperature field which is isotropic and leads to the well-known charmonium dissociation, the electromagnetic and rotational fields break down the radial symmetry, and the competition between strong interaction and electromagnetic and rotational interaction in the direction of Lorentz force makes the charmonia transit from an isotropic bound state of strong interaction with positive binding energy to an anisotropic bound state of electromagnetic and rotational interaction with negative binding energy. The transition seems possible to be realized in high energy nuclear collisions. 
\end{abstract}

\maketitle

Quarkonia have long been considered as a probe~\cite{matsui} of the new state of matter - quark-gluon plasma (QGP) which can be created in the early stage of high energy nuclear collisions. The cold and hot nuclear matter effects on quarkonium properties and the consequence in the final state of nuclear collisions are deeply investigated~\cite{andronic,dainese,aarts,fcc,zhao}. The study on quarkonia in medium is recently extended to including electromagnetic and rotational fields, since the strongest fields in nature can be generated in nuclear collisions~\cite{tuchin,marasinghe,deng,tuchin2,gubler,alford,guo,bonati,yoshida,liang,deng2,jiang,star}. Different from the electromagnetic fields which rapidly decay in time, the angular momentum conservation during the evolution of the collisions may make a more visible rotational effect on the final state. 

When the electromagnetic and rotational fields are strong enough, is it possible for a quarkonium state to transit from a bound state of strong interaction to a bound state of electromagnetic and rotational interaction? In this paper, we focus on the charmonium transition between the two kinds of bound states. As an effective theory to study bound states of heavy quarks, the non-relativistic and relativistic potential models, based on Schr\"odinger and Dirac equations, have been successfully used to describe quarkonium properties in vacuum~\cite{crater,shi} and medium~\cite{guo2,machado,zhao2,satz,karsch,du,chen} with the help from the lattice QCD simulated heavy quark potential~\cite{petreczky,burnier,lafferty}. We will take the Schr\"odinger equation to calculate the $c\bar c$ bound states in electromagnetic and rotational fields. 

The system under a rotational field can be equivalently regarded as a system at rest in a rotating frame, as has been discussed in Refs.\cite{Jiang:2016wvv,Chen2016,Liu:2018xip}. For a fermion system in a rotational field ${\bm \omega}$, the Lagrangian density in the rotating frame can be written as
\begin{equation}
\mathcal L = \bar\psi\left(i\gamma^\mu\partial_\mu+\gamma_0{\bm \omega}\cdot {\bm j}-m\right)\psi,
\end{equation}
where $m$ is the particle mass, ${\bm p}=-i{\bm \nabla}$ the momentum, and ${\bm l}={\bm x}\times {\bm p}$, ${\bm s}=-\gamma_0\gamma_5{\bm\gamma}/2=\text{diag}({\bm\sigma},{\bm\sigma})/2$ and ${\bm j}={\bm l}+{\bm s}$ are the orbital, spin and total angular momenta. With the Lagrangian density the Dirac equation reads 
\begin{equation}
\left(i\gamma^\mu\partial_\mu+\gamma_0{\bm \omega}\cdot{\bm j}-m\right)\psi=0
\end{equation}
which leads to the familiar one-body Hamiltonian density in the non-relativistic limit,  
\begin{equation}
\label{h0}
\mathcal H_{\omega i}={{\bm p}_i^2\over 2m}-{\bm \omega}\cdot{\bm j}_i
\end{equation}
with $i=c,\bar c$. For the two-body Schr\"odinger equation to describe the $c\bar c$ system with charge $q_c=-q_{\bar c}=q$ and mass $m_c=m_{\bar c}=m$, the Hamiltonian density reads
\begin{equation}
\label{h1}
H_\omega = \mathcal H_{\omega c} +\mathcal H_{\omega \bar c}+V_{c\bar c}+V_{ss},
\end{equation}
where we have included the confinement potential between $c$ and $\bar c$ which is usually chosen as the Cornell form $V_{c\bar c}(|{\bm x}_c-{\bm x}_{\bar c}|)=-\alpha/|{\bm x}_c-{\bm x}_{\bar c}|+\sigma |{\bm x}_c-{\bm x}_{\bar c}|$ and the spin-spin interaction $V_{ss}(|{\bm x}_c-{\bm x}_{\bar c}|,{\bm s}_c,{\bm s}_{\bar c})=\beta e^{-\gamma |{\bm x}_c-{\bm x}_{\bar c}|}{\bm s}_c\cdot {\bm s}_{\bar c}$. Both interactions are supported by the lattice QCD simulations~\cite{petreczky,burnier,lafferty,kawanai}, and the parameters $\alpha, \sigma, \beta$ and $\gamma$ can be fixed by fitting the charmonium properties in vacuum~\cite{zhao}. 

By introducing the total and relative coordinates and momenta ${\bm R}=({\bm x}_c+{\bm x}_{\bar c})/2, {\bm r}={\bm x}_c-{\bm x}_{\bar c},  {\bm P}={\bm p}_c+{\bm p}_{\bar c}$ and ${\bm p}=({\bm p}_c-{\bm p}_{\bar c})/2$, the total spin ${\bm s}={\bm s}_c+{\bm s}_{\bar c}$ and its projection $s_z=s_{cz}+s_{\bar cz}$, the total wave function can be separated into a center-of-mass part and a relative part $\Psi({\bm R},{\bm r},{\bm s},s_z)=\Theta({\bf R})\psi({\bm r},{\bm s},s_z)$. The relative wave function is governed by the Schr\"odinger equation
\begin{equation}
\label{s1}
\left({{\bm p}^2\over m} -{\bm \omega}\cdot({\bm l}+{\bm s})+V_{c\bar c}+V_{ss}\right)\psi=\epsilon\psi
\end{equation}
with the relative orbital angular momentum ${\bm l}={\bm r}\times{\bm p}$ and binding energy $\epsilon$. The relative equation characterizes the inner structure of the $c\bar c$ state. While the total wave function can be factorized as a center-of-mass and a relative part, the total momentum ${\bm P}$ is not conserved during the evolution of the system, $[{\bm P}, H_\omega]\neq 0$, due to the global rotation, and the center-of-mass motion is not a plane wave.

For fermions in a gauge potential $A_\mu(x)$, if one defines the electromagnetic fields ${\bm E}$ and ${\bm B}$ via $A_\mu = ({\bm E}\cdot{\bm x},({\bm B}\times{\bm x})/2)$, the one- and two-body Hamiltonian densities read   
\begin{eqnarray}
\label{h2}
\mathcal H_{Ei} &=& {({\bm p}_i-{q_i\over 2}({\bm B}\times{\bm x}_i))^2\over 2m}-q_i{\bm E}\cdot{\bm x}_i-{q_i\over m}{\bm B}\cdot{\bm s}_i,\nonumber\\
H_E &=& \mathcal H_{Ec}+\mathcal H_{E\bar c}+V_{c\bar c}+V_{ss}.
\end{eqnarray}
Alford and Strickland~\cite{alford} studied systematically the charmonium properties in electromagnetic fields. Different from the rotational field, the pseudo-momentum ${\bm P}_{ps}={\bm P}+q({\bm B}\times {\bm r})/2$ is conserved in electromagnetic fields, $[{\bm P}_{ps},H_E]=0$. Therefore, the total wave function can be factorized as $\Psi({\bm R},{\bm r},{\bm s},s_z)=\Theta({\bm R},{\bm r})\psi({\bm r},{\bm s},s_z)$ with a modified plane wave $\Theta({\bm R},{\bm r})=e^{i{\bm R}\cdot({\bm P}_{ps}-q({\bm B}\times {\bm r})/2)}$, and the relative part is controlled by the Schr\"odinger equation
\begin{eqnarray}
\label{s2}
&& \bigg({{\bm p}^2\over m}+{q^2({\bm B}\times{\bm r})^2-2q {\bm P}_{ps}\cdot({\bm B}\times{\bm r}) \over 4m}-q{\bm E}\cdot{\bm r}\nonumber\\
&& -{q\over m}{\bm B}\cdot({\bm s}_c-{\bm s}_{\bar c})+V_c +V_{ss}\bigg)\psi=\epsilon\psi.
\end{eqnarray}

We now consider both the rotational and electromagnetic fields. In this case we face the problem of which frame we introduce the electromagnetic fields in. If one defines the gauge potential $A_\mu = ({\bm E}\cdot{\bm x},({\bm B}\times{\bm x})/2)$ in the rotating frame~\cite{Matsuoprb,Matsuoprl,anandan}, we replace the momentum ${\bm p}$ in the one-body Hamiltonian (\ref{h0}) which is defined in the rotating frame too by ${\bm p}-q({\bm B}\times {\bm x})/2$, and the one- and two-body Hamiltonian densities read
\begin{eqnarray}
\label{h3}
\overline{\mathcal H}_i &=& \mathcal H_{Ei}-{\bm \omega}\cdot\left({\bm x}_i\times({\bm p}_i-{q_i\over 2}({\bm B}\times{\bm x}_i))+{\bm s}_i\right),\nonumber\\
\overline H &=& \overline{\mathcal H}_c+\overline{\mathcal H}_{\bar c}+V_{c\bar c}+V_{ss}\nonumber\\
&=&H_E-{\bm \omega}\cdot\left({\bm j}_c+{\bm j}_{\bar c}\right)\nonumber\\
&&+{q\over 2}{\bm \omega}\cdot\left({\bm x}_c\times({\bm B}\times{\bm x}_c)-{\bm x}_{\bar c}\times({\bm B}\times{\bm x}_{\bar c})\right).
\end{eqnarray}
It is clear that the last term is a mixing between the rotational field ${\bm \omega}$ and the magnetic field ${\bm B}$. 

However, in heavy ion collisions people usually measure or calculate the electromagnetic fields in the laboratory frame. In this case, one should make transformation for the electromagnetic potential between the local rest and laboratory frames~\cite{Chen2016,Liu:2018xip}. From the Lagrangian density in the rotating frame 
\begin{equation}
\mathcal L = \sqrt{-g}\bar\psi\left(i\bar\gamma^\mu (\partial_\mu-\Gamma_{\mu})-q\bar\gamma^\mu\bar A_\mu-m\right)\psi,
\end{equation} 
where $\Gamma_{\mu}$ is the affine connection, $\bar\gamma_\mu$ the gamma matrix satisfying $[\bar \gamma_{\mu},\bar\gamma_{\nu}]=2g_{\mu\nu}$, and $\bar A_\mu(x)$ the electromagnetic potential. We transform the vectors $\bar\gamma_\mu$ and $\bar A_\mu(x)$ in the rotating frame into $\gamma_\alpha$ and $A_\alpha(x)$ in the local rest frame with tetrad $e^{\alpha\ }_{\ \mu}$ satisfying $g_{\mu\nu}=e^{\ \alpha}_{\mu\ }\eta_{\alpha\beta}e^{\beta\ }_{\ \nu}$. Then we connect the potential $A_\alpha(x)$ to the measured or calculated $A'_a(x')$ in the laboratory frame through the transformation $A_{\alpha}(x)=A'_a(\Lambda^{a\ }_{\ \mu}(x)e^{\mu\ }_{\ \alpha}+\Lambda^{a\ }_{\ \nu,\mu}(x)e^{\mu}_{\alpha} x^{\nu})$, where $\Lambda^{a\ }_{\ \mu}(x)$ is an arbitrary local coordinate transformation. If we still define the electromagnetic fields ${\bm E}'(x')$ and ${\bm B}'(x')$ via $A_a'=({\bm E}'\cdot{\bm x}', -({\bm B}'\times{\bm x}')/2)$ and take all the three external fields $({\bm E}, {\bm B}, {\bm \omega})$ in the same direction, the Lagrangian density becomes finally  
\begin{equation}
\mathcal L=\bar\psi\left(i\gamma^{\mu}\partial_{\mu}+\gamma^0({\bm\omega}\cdot{\bm j})-q{\gamma}^\mu A_\mu-m\right)\psi
\end{equation}
in the rotating frame. Considering the non-relativistic limit of the Dirac equation, the one- and two-body Hamiltonian densities become 
\begin{eqnarray}
\label{h4}
\mathcal H_i &=& \mathcal H_{Ei}-{\bm \omega}\cdot{\bm j}_i,\nonumber\\
H &=& \mathcal H_c+\mathcal H_{\bar c}+V_{c\bar c}+V_{ss}\nonumber\\
&=& H_E-{\bm \omega}\cdot\left({\bm j}_c+{\bm j}_{\bar c}\right).
\end{eqnarray}
The mixing between the rotational field and magnetic field disappears in this case. It is necessary to emphasize that, the disappearance of the mixing holds only under the condition of parallel external fields, which is approximately the case in heavy ion collisions. For a general case with different directions of the three external fields, there will be mixing terms among them in the Hamiltonian density. After the transformation from individual variables to center-of-mass and relative variables, the total angular momentum is separated into the center-of-mass and relative parts ${\bm j}_c+{\bm j}_{\bar c}={\bm L}+{\bm l}+{\bm s}$ with the total orbital angular momentum ${\bm L}={\bm R}\times{\bm P}$. Different from the case with only rotational field or only electromagnetic fields where the total wave function can always be separated into a center-of-mass and a relative part, it becomes impossible to factorize the $c\bar c$ motion in rotational and electromagnetic fields. This is true for both the Hamiltonian densities $\overline H$ and $H$.        

Since there is no longer a relative equation to directly study the bound state properties, we have to use a perturbative method to approximately solve the total Sch\"odinger equation $H\Psi=E\Psi$. Considering the fact that, in nuclear collisions at RHIC and LHC the strength of the magnetic field $eB\simeq 70 m_\pi^2$ is much larger than the rotational field $m\omega\simeq m_\pi^2$ for charm quarks, we can choose the electromagnetic coupling as the main part of the total interaction and take the rotational coupling as a perturbation. Therefore, we separate the Hamiltonian (\ref{h4}) into a main and a perturbative part,  
\begin{equation}
\label{ht3}
H = H_E + H'
\end{equation}
with
\begin{eqnarray}
\label{hp}
H' &=& H'_c + H'_r,\nonumber\\
H'_c &=& -{\bm\omega}\cdot {\bm L}_{ps},\nonumber\\
H'_r &=& - {\bm \omega}\cdot({\bm l}+{\bm s})+ {q\over 2}{\bm\omega}\cdot \left({\bm R}\times({\bm B}\times{\bm r})\right),
\end{eqnarray}
where $H'_c$ and $H'_r$ are the corrections from the rotation to the center-of-mass motion and relative motion. For the main part $H_E$, ${\bm P}_{ps}$ is the conserved momentum, we must keep it in the perturbation. This is the reason why we use ${\bm L}_{ps}={\bm R}\times {\bm P}_{ps}$ in the perturbation $H'_c$, and this also leads to the mixing between the rotation and electromagnetic fields in the perturbation $H'_r$. Note that, the mixing here is due to the perturbation used.  

The contribution from the perturbation $H'$ can systematically be calculated through the standard method in quantum mechanics. To the first order, the binding energy $\epsilon$ and relative wave function $\psi$ are 
\begin{eqnarray}
\epsilon_n &=& \epsilon_n^{(0)}+\langle\psi_n^{(0)}|H'_r|\psi_n^{(0)}\rangle,\nonumber\\
{\psi}_n &=& \psi_n^{(0)}+\sum_{m\not = n}{\langle \psi_m^{(0)}|H'_r|\psi_n^{(0)}\rangle\over \epsilon_m^{(0)}-\epsilon_n^{(0)}}\psi_m^{(0)},
\end{eqnarray}
where $\epsilon_n^{(0)}$ and $\psi_n^{(0)}$ of the $c\bar c$ bound state with quantum number $n$ are controlled by the Schr\"odinger equation in electromagnetic fields. The center-of-mass coordinate ${\bf R}$ in the perturbation $H'_r$ should be considered as the average one, $\langle {\bf R} \rangle = \int _0^{{\bf R}_{max}}|\Theta|^2{\bf R}d^3{\bf R}/\int _0^{{\bf R}_{max}}|\Theta|^2d^3{\bf R}$, where ${\bf R}_{max}$ is the size of the system. For a constant rotation, to guarantee the law of causality, the size is under the constraint of $|R_{max}\omega|\leq 1$. For $\omega\sim 0.1 m_\pi$ at RHIC and LHC, there is $R_{max}\leq 15$ fm which is about the maximum size of the QGP created in the collisions.

We first analyze the possible transition from the bound state of strong interaction to the bound state of electromagnetic and rotational interaction. From the Hamiltonian (\ref{ht3}) the effective potential $V({\bm r}|{\bm E},{\bm B}, {\bm \omega},{\bm P}_{ps},\langle {\bm R}\rangle)$ between $c$ and $\bar c$ is
\begin{eqnarray}
V &=& V_{c\bar c}+V_{ss}+{q^2({\bm B}\times{\bm r})^2-2q{\bm P}_{ps}\cdot({\bm B}\times{\bm r})\over 4m}\nonumber\\
&&-q{\bm E}\cdot{\bm r}-{q\over m}{\bm B}\cdot({\bm s}_c-{\bm s}_{\bar c})+H'_r.
\end{eqnarray}
The first two terms $V_{c\bar c}$ and $V_{ss}$ are strong interactions with radial symmetry, and the other terms are electromagnetic and rotational interactions which break down the radial symmetry and therefore will enhance or reduce the strong interaction in different directions. To be specific, we consider the electromagnetic and rotational fields created in heavy ion collisions~\cite{deng,deng2}. If we take the $y$-axis as the beam line of the collisions, the maximum magnetic field ${\bm B}$ and rotational field ${\bm \omega}$ are along the direction of ${\bf e}_z$. In the central rapidity region of the collisions, the electric field is much smaller than the magnetic field, and only the $z$-component is relatively sizeable. Under this consideration the potential becomes $V({\bm r}|E{\bf e}_z,B{\bf e}_z,\omega{\bf e}_z,{\bm P}_{ps}^\perp,\langle {\bm R}_\perp\rangle)$, where $\langle{\bm R}_\perp\rangle$ and ${\bm P}_{ps}^\perp$ are the transverse coordinate and momentum. The enhancement or cancellation between the strong interaction and electromagnetic and rotational interactions depends strongly on the directions of ${\bf P}_{ps}^\perp$ and $\langle {\bf R}_\perp\rangle$. It is easy to see that the maximum and minimum potentials are $V_\pm({\bm r}|E{\bf e}_z,B{\bf e}_z,\omega{\bf e}_z,P_{ps}^\perp{\bf e}_x,\pm\langle R_\perp\rangle{\bf e}_y)$. The potential with any other ${\bm P}_{ps}^\perp$ and $\langle {\bm R}_\perp\rangle$ is between the two limits. 
\begin{figure}[!htb]
	\includegraphics[width=0.45\textwidth]{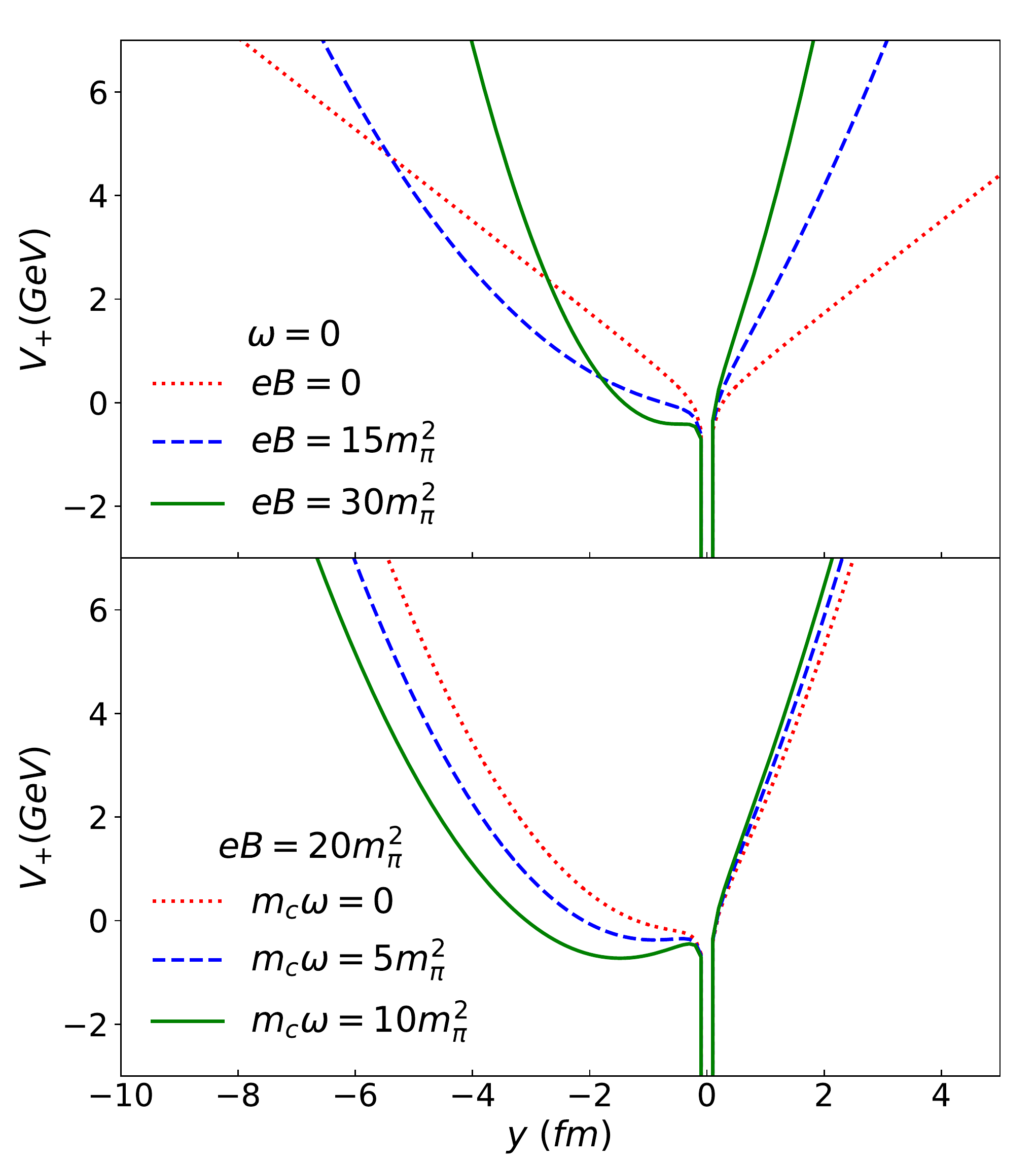} 
	\caption{The spin independent part of the maximum charmonium potential $V_+$ as a function of $y$ at $x=z=0$ in magnetic field (upper panel) and both magnetic and rotational fields (lower panel). The transverse momentum and coordinate are taken as $P_{ps}^\perp =2.5$ GeV and $\langle R_\perp\rangle =4/3$ fm. }
	\label{fig1}
\end{figure}

The spin independent part of the maximum potential $V_+$ as a function of $y$ at $x=z=0$ is demonstrated in Fig.~\ref{fig1}. The upper panel shows the pure magnetic effect (the electric effect $-qEz$ disappears at $z=0$). The parameters $\alpha$, $\sigma$, $m$ and $P_{ps}^\perp$ are taken as the usually used values~\cite{zhao} $\alpha=0.312,\ \sigma=0.174$ (GeV)$^2,\ m=1.29$ GeV and $P_{ps}^\perp=2.5$ GeV. The singularity at $y=0$ comes from the Coulomb potential $-\alpha/|y|$. Without magnetic field the potential is a symmetric function of $y$, but the symmetry is broken when the magnetic field is turned on. Around the origin the potential is enhanced at $y>0$ but suppressed at $y<0$ by the magnetic field. When the field is strong enough, a new potential well forms, and the $c\bar c$ pair transit from the bound state of strong interaction to the bound state of magnetic interaction. The location of the new well is controlled by the condition 
\begin{equation}
dV_+/dy =0.
\end{equation}

When the rotational field is switched on, the rotation dependence of the potential is shown in the lower panel of Fig.\ref{fig1} with fixed magnetic field and charmonium transverse coordinate. The enhancement at $y>0$ and cancellation at $y<0$ become now more visible. It is clear that the rotational field deepens the new potential well and accelerates the transition between the two kinds of bound states.  

We now perturbatively solve the charmonium binding energy and wave function. Considering the spin interaction, the relative wave function $\psi^{(0)}({\bm r},s_c,s_{\bar c})$ can not be factorized as a spatial part and a spin part in general case. We use the four independent spin states $|s,s_z\rangle$: the spin singlet state $|S\rangle = |0,0\rangle$ and triplet states $|T_0\rangle = |1,0\rangle$ and $|T_\pm\rangle = |1,\pm 1\rangle$. They satisfy the relations ${\bm B}\cdot({\bm s}_c-{\bf s}_{\bar c}) |T_\pm\rangle = 0,\ {\bm B}\cdot({\bm s}_c-{\bm s}_{\bar c}) |T_0\rangle = B|S\rangle,\ {\bm B}\cdot({\bm s}_c-{\bm s}_{\bar c}) |S\rangle = B|T_0\rangle,\ {\bm \omega}\cdot{\bm s}|T_\pm\rangle = \pm\omega|T_\pm\rangle,\ {\bm \omega}\cdot{\bm s}|T_0\rangle = 0,\ {\bm \omega}\cdot{\bm s}|S\rangle = 0,\ {\bm s}_c\cdot{\bm s}_{\bar c}|T_\pm\rangle = {1\over 4}|T_\pm\rangle,\ {\bm s}_c\cdot{\bm s}_{\bar c}|T_0\rangle = {1\over 4}|T_0\rangle$ and ${\bm s}_c\cdot{\bm s}_{\bar c}|S\rangle = -{3\over 4}|S\rangle$. While the coupling between spin and magnetic field keeps the triplet states $|T_\pm\rangle$ as the eigenstates of the Hamiltonian $H_E$, it leads to a mixing between $|S\rangle$ and $|T_0\rangle$~\cite{Cho2014}. For the coupling between spin and rotational field, it does not make any mixing among the spin states, but creates a energy gap $\sim 2\omega$ between the two triplet states $|T_\pm\rangle$. The spin-spin interaction splits the singlet and triplet states.

The relative equation with only strong interaction can be separated into a radial part and an angular part. The binding energy is determined by the radial equation, and the solution of the angular part is the spherical harmonic function $Y_{lm}(\theta,\phi)$. Considering the direction dependence of electromagnetic and rotational interactions, the potential between the quark and anti-quark is no longer a central one, a usual way to solve the relative equation with electromagnetic interaction is to expand the wave function $\psi^{(0)}$ in terms of the complete and orthogonal spherical harmonic functions,
\begin{eqnarray}
\label{expansion}
{r\psi_T^{(0)\pm}({\bf r}, s, s_z)} &=&\sum_{lm}a_{lm}^\pm u_{lm}^\pm(r)Y_{lm}(\theta,\phi)|T_\pm\rangle,\nonumber\\
{r\psi_{T,S}^{(0)0}({\bf r}, s, s_z)} &=& \sum_{lm}\big[a_{lm}^0 u_{lm}^0(r)Y_{lm}(\theta,\phi)|T_0\rangle\nonumber\\
&& +a_{lm}^Su_{lm}^S(r)Y_{lm}(\theta,\phi)|S\rangle\big],
\end{eqnarray}
where $a_{lm}^\pm, a_{lm}^0$ and $a_{lm}^S$ are the probability amplitudes for spin triplet and singlet states. By substituting the expansion into the relative equation, we derive the wave equations controlling the radial components $u_{lm}^\pm(r), u_{lm}^0(r)$ and $u_{lm}^S(r)$.

We apply the inverse power method~\cite{crater} to numerically solving the radial equations for the charmonium ground states $J/\psi$ and $\eta_c$. The spin triplet states $J/\psi_\pm$ and $J/\psi_0$ cannot be distinguished in vacuum, they are all called $J/\psi$. In electromagnetic and rotational fields, their spin-spin interaction and spin-field interaction are different, and $J/\psi_0$ is coupled with the spin singlet state $\eta_c$. By fitting the experimentally observed charmonium masses in vacuum~\cite{tanabashi}, we fix the parameters in the spin sector of the potential model: $\beta=1.982$ GeV and $\gamma = 2.06$ GeV. With the known wave function $\psi^{(0)}$ and binding energy $\epsilon^{(0)}$ in pure electromagnetic fields, we turn to consider the correction from the rotational effect. We first check the convergence of the perturbative expansion by calculating the relative correction  
\begin{equation}
{\langle H' \rangle\over E_E}= \frac{\langle\psi^{(0)}|\omega(\langle R_{\perp}\rangle P_{ps}-l -s +{1\over 2} qB \langle R_{\perp}\rangle y)|\psi^{(0)}\rangle}{\epsilon^{(0)}+\frac{P_{ps}^2}{4m}}.
\end{equation}
Taking the above mentioned parameters, the correction for $J/\psi_0$ is $9.4\%$ at $\epsilon^{(0)}=0$. Note that the correction at $\epsilon^{(0)}=0$ is the maximum correction and we have $\langle H'\rangle/E_E \leq 9.4\%$ in general case. This means a fast convergence of the expansion. 

\begin{figure}[!htb]
\includegraphics[width=0.45\textwidth]{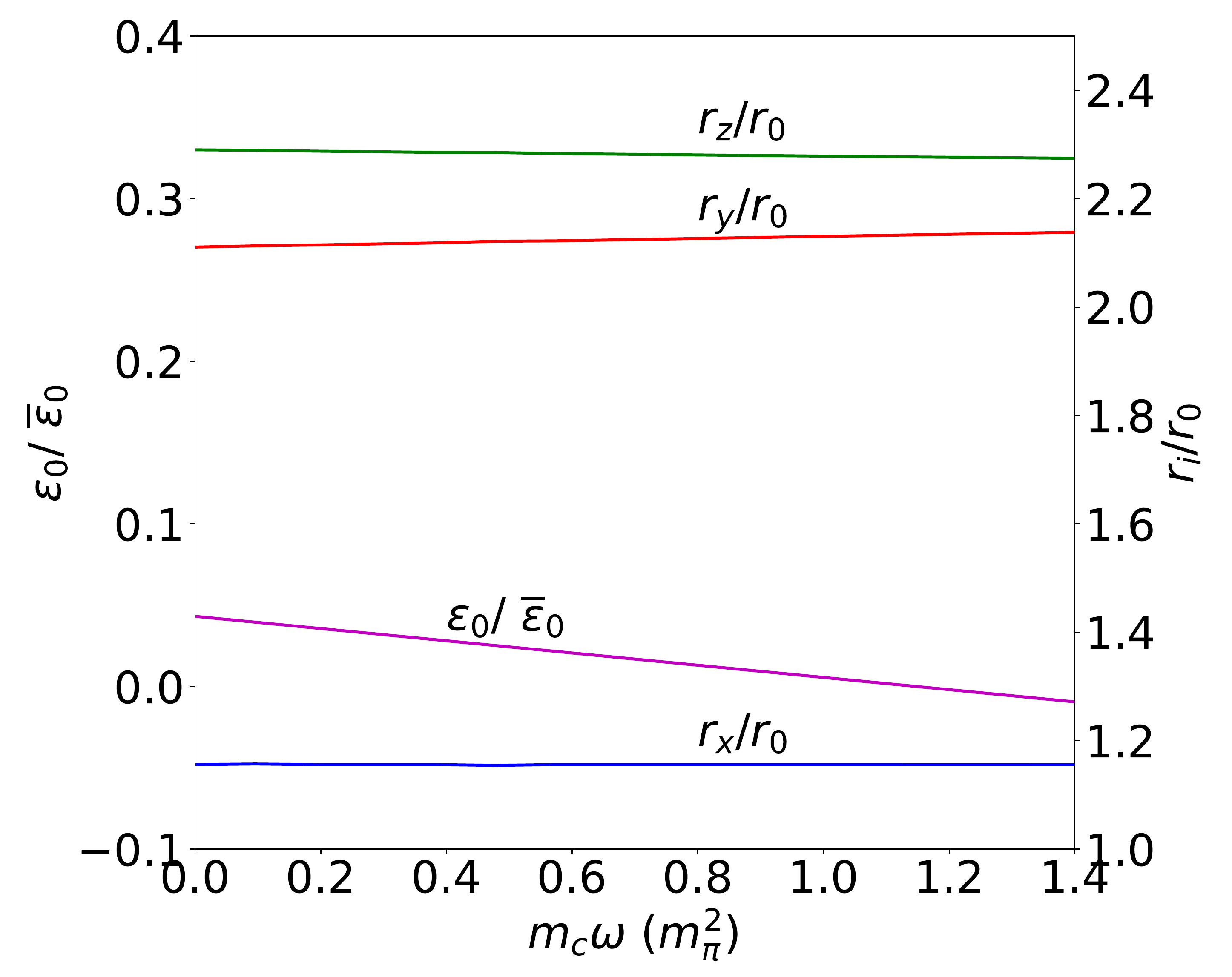}
\caption{The rotation dependence of the binding energy $\epsilon_0$ and root-mean-squared radii $r_x,\ r_y$ and $r_z$ for $J/\psi_0$ with maximum potential $V_+$. The parameters are fixed to be $eE=12\ m_\pi^2, eB=28\ m_\pi^2, P_{ps}^\perp$=2.5 GeV and $\langle R_\perp\rangle$=2/3 fm, and the scaled parameters $\bar\epsilon_0$ and $r_0$ are the $J/\psi$ binding energy and averaged radius in vacuum. }
\label{fig2}
\end{figure}
The binding energy $\epsilon_0$ and root-mean-squared radii $r_i\ (i=x,y,z)$ for the ground state of $J/\psi_0$ in electromagnetic and rotational fields are shown in Fig.\ref{fig2} in the case with maximum interaction potential. To focus on the rotational effect, the electromagnetic fields $E$ and $B$ are fixed. In the beginning at $\omega=0$, the binding energy $\epsilon_0$ is already reduced to only $4.5\%$ of its vacuum value $\bar \epsilon_0$ by the electromagnetic effect. With increasing rotation, the potential in the direction of Lorentz force is continuously suppressed, the binding energy drops down monotonously. At the transition point $m\omega = 1.15\ m_\pi^2$, the binding energy approaches to zero, and the bound state of strong interaction vanishes. Beyond the transition point, the potential in the direction of the Lorentz force becomes negative, and the $c\bar c$ pair is in the bound state of electromagnetic and rotational interaction with negative binding energy. For the charmonium shape, we consider two quantities: the fluctuation $\langle{\bm r}\rangle=\int d^3{\bm r}{\bm r}|\psi({\bm r})|^2$ and the three radii $r_i=[\int d^3{\bm r}r_i^2|\psi({\bm r})|^2]^{1/2},\ i=x,y,z$. The former indicates the degree of the radial symmetry breaking by the electromagnetic and rotational fields, and the latter describes the charmonium size in different directions. For a central potential, the wave function is radial symmetric with $\langle{\bm r}\rangle=0$. In electromagnetic and rotational fields, the wave function becomes anisotropic. From the interaction potential, the electric force stretches the $c$ and $\bar c$ in the $z$-direction, the Lorentz force broadens the wave function along the $y$-direction and leads to a fluctuation $\langle y\rangle <0$, and the force coming from the oscillation potential reduces the size in the $x$- and $y-$directions. Since the wave function cannot be factorized, the sizes shown in Fig.\ref{fig2} are controlled by the competition among all the strong, electromagnetic and rotational interactions. As we pointed out above, the electromagnetic field is stronger than the rotational field in high-energy nuclear collisions, $eB>m\omega$, the change in charmonium structure is mainly due to the electromagnetic field, and the rotation is only a perturbation. This can clearly be seen in Fig.\ref{fig2}: The rotation dependence of both the binding energy and root-mean-squared radii is weak.

At this point we want to emphasize the difference from the charmonia and even bottomonia dissociation in hot medium. One difference is the anisotropy of electromagnetic and rotational interaction. The dissociation at high temperature is isotropic, but the transition in strong electromagnetic and rotational fields happens only in the direction around the Lorentz force. The other difference is the broadening of the relative wave function. The thermal motion suppresses the long-distance part but keeps the short-distance part of the strong interaction~\cite{petreczky,burnier,lafferty}, which leads to a tremendously broadening of the charmonium wave function, while the Lorentz force and Coriolis force change both the long- and short-range interaction, which will not sizeably modify the charmonium distribution. This is the reason why the averaged charmonium radius approaches to infinity at the dissociation point in hot medium but the averaged radii do not change tremendously around the transition point.         
\begin{figure}[!htb]
\includegraphics[width=0.45\textwidth]{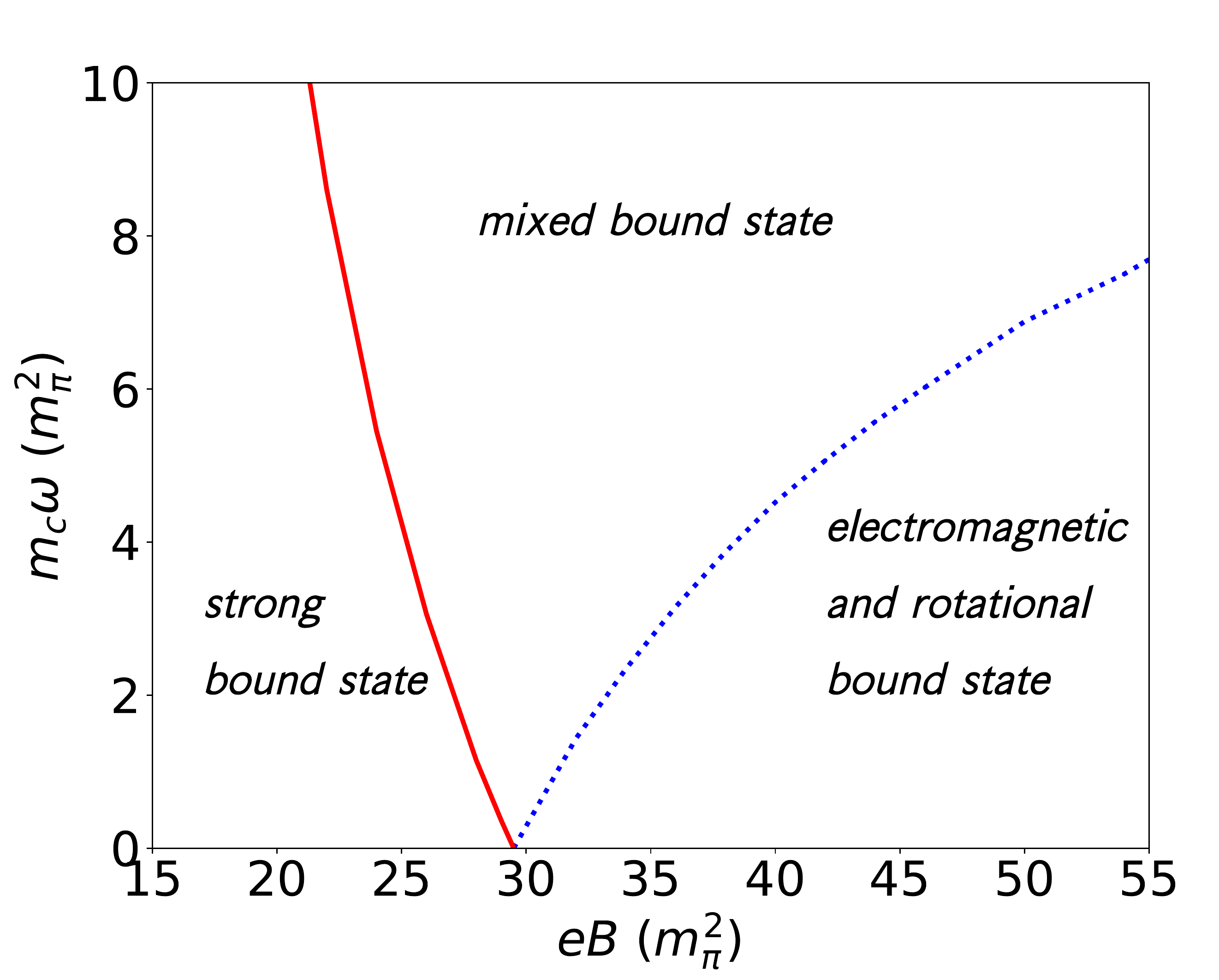}
\caption{The $J/\psi_0$ structure in rotation-magnetic field plane $m\omega - eB$ at fixed electric field $eE=12\ m_\pi^2$. The transverse momentum and coordinate of $J/\psi_0$ are taken as $P_{ps}^\perp=2.5$ GeV and $\langle R_\perp\rangle =2/3$ fm.}
\label{fig3}
\end{figure}

We now turn to calculating the $J/\psi_0$ transition line defined by $\epsilon_0=0$ in $m\omega - eB$ plane at fixed electric field $eE=12\ m_\pi^2$. The result is shown in Fig.\ref{fig3}. When the magnetic field is too weak, the rotation under the constraint of the law of causality cannot trigger the charmonium transition, the $c\bar c$ pair is in the bound state of strong interaction. With increasing magnetic field, the transition from strongly to electromagnetically and rotationally interacting bound state happens with the maximum potential $V_+$. The transition rotation drops down fast, see the solid line. At the magnetic field $eB\simeq 29.5\ m_\pi^2$, the transition takes place without the help from the rotation. The left side of the solid line is the region of $c\bar c$ bound state of strong interaction with binding energy $\epsilon_0 > 0$, and the right side of the line is the region of the bound state of electromagnetic and rotational interaction with $\epsilon_0 <0$. In the other limit with the minimum potential $V_-$, since the rotation reduces the electromagnetic effect, the transition rotation increases with magnetic field, see the dashed line. Again, the left side and right side of the line are respectively strong interaction induced bound state and electromagnetic and rotational interaction induced bound state. The transition line with any other potential $V$ is in between the two lines. Therefore, the region between the two lines is a mixed phase with both strong and electromagnetic and rotational bound states. Note that, the condition to form an electromagnetic and rotational interaction controlled charmonium state, $eB\gtrsim 29.5\ m_\pi^2$ and $m\omega \lesssim 8\ m_\pi^2$, seems possible to be realized in high energy nuclear collisions.        

We investigated in a potential model the charmonium transition from the bound state of strong interaction to the bound state of electromagnetic and rotational interaction. When both the electromagnetic and rotational fields are turned on, the Hamiltonian of the $c\bar c$ pair depends on the frame where the gauge fields are introduced via minimum coupling principle. Taking the fields created in high-energy nuclear collisions, we calculated the interaction potential between the $c$ and $\bar c$. With increasing fields, the strong interaction which confines the two heavy quarks in vacuum is gradually canceled by the electromagnetic and rotational interaction, and the transition from strongly to electromagnetically and rotationally interacting bound state happens along the direction of the Lorentz force. Very different from the charmonium dissociation in hot medium, the transition is anisotropic and the charmonium size does not change significantly during the transition process. The condition of the transition can be realized in high-energy nuclear collisions.         

{\bf Acknowledgement:} We thank Drs. Shuzhe Shi and Partha Bagchi for helpful discussions in the beginning of the work. The work is supported by Guangdong Major Project of Basic and Applied Basic Research No. 2020B0301030008 and the NSFC under grant Nos. 11890712 and 12075129.


\begin{thebibliography}{20}

\bibitem{matsui} T.Matsui and H.Satz, Phys. Lett. {\bf B178}, 416(1986).

\bibitem{andronic}
A.Andronic {\it et al.}, Eur. Phys. J. {\bf C76}, 107(2016). 

\bibitem{dainese} 
A.Dainese {\it et al.}, CERN Yellow Rep., 635(2017).  

\bibitem{aarts} 
G.Aarts {\it et al.}, Eur. Phys. J. {\bf A53}, 93(2017).

\bibitem{fcc} 
A.Abada {\it et al.} [FCC Collaboration], Eur. Phys. J. {\bf C79}, 474(2019).

\bibitem{zhao}
J.~Zhao, K.~Zhou, S.~Chen and P.~Zhuang, Prog. Part. Nucl. Phys. {\bf 114}, 103801(2020).

\bibitem{tuchin}
K.Tuchin, Phys. Lett. {\bf B705}, 482(2011). 

\bibitem{marasinghe}
K.Marasinghe and K.Tuchin, Phys. Rev. {\bf C84}, 044908(2011). 

\bibitem{deng} 
W.Deng and X.Huang, Phys. Rev. {\bf C85}, 044907(2012).

\bibitem{tuchin2} 
K.Tuchin, Adv. High Energy Phys. {\bf 2013}, 490495(2013).

\bibitem{gubler}
P.Gubler, K.Hattori, S.H.Lee, M.Oka, S.Ozaki and K.Suzuki, Phys. Rev. {\bf D93}, 054026(2016).

\bibitem{alford} 
J.Alford and M.Strickland, Phys. Rev. {\bf D88}, 105017(2013).

\bibitem{guo}
X.Guo, S.Shi, N.Xu, Z.Xu and P.Zhuang, Phys. Lett. {\bf B751}, 215(2015).

\bibitem{bonati} 
C.Bonati, M.D'Elia and A.Rucci, Phys. Rev. {\bf D92}, 054014(2015).

\bibitem{yoshida}
T.Yoshida and K.Suzuki, Phys. Rev. {\bf D94}, 074043(2016). 

\bibitem{liang} 
Z.Liang and X.Wang, Phys. Rev. Lett. {\bf 94}, 102301(2005), Erratum: Phys. Rev. Lett. {\bf 96}, 039901(2006).

\bibitem{deng2} 
W.Deng and X.Huang, Phys. Rev. {\bf C93}, 064907(2016).

\bibitem{jiang} 
Y.Jiang, Z.Lin and J.Liao, Phys. Rev. {\bf C94}, 044910(2016), Erratum: Phys. Rev. {\bf C95}, 049904(2017).

\bibitem{star}
L.Adamczyk {\it et al.} [STAR Collaboration], Nature {\bf 548}, 62(2017).

\bibitem{crater} 
H.Crater and P.Van Alstine, Phys. Rev. {\bf D36}, 3007(1987).

\bibitem{shi}
S.Shi, J.Zhao and P.Zhuang, Chin. Phys. {\bf C44}, 084101(2020)  

\bibitem{satz} H.Satz, J. Phys. {\bf G32}, R25(2006).

\bibitem{karsch}
F.Karsch, D.Kharzeev and H.Satz, Phys. Lett. {\bf B637}, 75(2006). 

\bibitem{guo2}
X.Guo, S.Shi and P.Zhuang, Phys. Lett. {\bf B718}, 143(2012). 

\bibitem{machado}
C.Machado, F.Navarra, E.de Oliveira, J.Noronha and M.Strickland, Phys. Rev. {\bf D88}, 034009(2013).

\bibitem{chen}
B.Chen, Y.Liu, K.Zhou and P.Zhuang, Phys. Lett. {\bf B726}, 725(2013). 

\bibitem{du}
X.Du and R.Rapp, Nucl. Phys. {\bf A943}, 147(2015). 

\bibitem{zhao2} 
J.Zhao and P.Zhuang, Few Body Syst. {\bf 58}, 100(2017).

\bibitem{petreczky} 
P.Petreczky, J. Phys. {\bf G37}, 094009(2010).

\bibitem{burnier}
Y.Burnier, O.Kaczmarek and A.Rothkopf, Phys. Rev. Lett. {\bf 114}, 082001(2015).

\bibitem{lafferty}
D.Lafferty and A.Rothkopf, Phys. Rev. {\bf D101}, 056010(2020).

\bibitem{Jiang:2016wvv}
Y.Jiang and J.Liao, Phys. Rev. Lett. {\bf 117}, 192302(2016).

\bibitem{Chen2016}
H.Chen, K.Fukushima, X.Huang and K.Mameda, Phys. Rev. {\bf D93}, 104052(2016). 

\bibitem{Liu:2018xip}
Y.C.Liu, L.L.Gao, K.Mameda and X.G.Huang, Phys. Rev. {\bf D99}, 085014(2019). 

\bibitem{kawanai} 
T.Kawanai and S.Sasaki, Phys. Rev. {\bf D85}, 091503(2012).

\bibitem{Matsuoprb} 
M.Matsuo, J.Ieda, E.Saitoh and S.Maekawa, Phys. Rev. {\bf B84}, 104410(2011).

\bibitem{Matsuoprl} 
M.Matsuo, J.Ieda, E.Saitoh and S.Maekawa, Phys. Rev. Lett. {\bf 106}, 076601(2011).

\bibitem{anandan} 
J.Anandan and J.Suzuki, arXiv: quant-ph/0305081.

\bibitem{Cho2014}
S.Cho, K.Hattori, S.Lee, K.Morita and S.Ozaki, Phys. Rev. Lett. {\bf 113}, 172301(2014). 

\bibitem{tanabashi} 
M.Tanabashi {\it et al.} [Particle Data Group], Phys. Rev. {\bf D98}, 030001(2018).

\end{thebibliography}
\end{document}